# Dynamic conformal arcs for lung stereotactic body radiation therapy: a comparison with volumetric modulated arc therapy

May 2019

Rasmus Bokrantz[1,*], Minna Wedenberg[1], Peter Sandwall[2]

1. RaySearch Laboratories, Stockholm, Sweden.
2. Department of Radiation Oncology, OhioHealth, Mansfield, OH, United States of America.
*   E-mail: rasmus.bokrantz@raysearchlabs.com

**ABSTRACT**

This study evaluates dynamic conformal arc (DCA) therapy as an alternative to volumetric modulated arc therapy (VMAT) for stereotactic body radiation therapy (SBRT) of lung cancer. The rationale for DCA is lower geometric complexity and hence reduced risk for interplay errors induced by respiratory motion. Forward planned DCA and inverse planned DCA based on segment-weight optimization were compared to VMAT for single arc treatments of three lung patients. Analysis of dose-volume histograms and clinical goal fulfillment revealed that DCA can generate satisfactory and near equivalent dosimetric quality to VMAT, except for complex tumor geometries. Segment-weight optimized DCA provided spatial dose distributions qualitatively similar to those for VMAT. Our results show that DCA, and particularly segment-weight optimized DCA, is an attractive alternative to VMAT for lung SBRT treatments, where errors from intrafraction motion cannot be expected to average out over the course of the treatment.

## 1. INTRODUCTION

Stereotactic body radiation therapy (SBRT) is standard of care for inoperable non-small cell lung cancer (NSCLC) (Ettinger, et al. 2017). In the United States, SBRT for NSCLC is delivered in 1-5 fractions and with up to 10 fractions internationally, both using biological effective doses in excess of 100 Gy (Videtic, et al. 2017). Radiation dose delivery to moving targets, such as lung tumors, has been a fundamental challenge in radiation oncology. Traditional approaches to account for motion have entailed expansion of the gross or clinical target volume to include the entire range of motion; defined as the internal target volume (ITV) (Chavaudra and Bridier, 2001). Several other devices and strategies have been developed to manage and minimize the effects of respiratory motion, including compression and breath hold devices (Keall, et al. 2006). A recent advance is the development of robust radiotherapy plans, where the uncertainty in the target location is parameterized in the optimization (Unkelbach, et al. 2018).

Currently, the most common approach for lung SBRT is treatment to an ITV using volumetric modulated arc therapy (VMAT), where dose is delivered during an arc with simultaneous dynamic motion of the multi-leaf collimator (MLC) leaves (Palma, et al. 2010). Current optimization methods do not constrain leaf motion to prevent occlusion of the target. This form of treatment delivery is susceptible to dosimetric errors from unexpected interplay between organ motion and MLC leaf motion—a phenomenon termed the interplay effect (Jiang, et al. 2003 and Seco, et al. 2007). The interplay effect can create significant dosimetric deviations greater than 20%; however, the effect averages out over traditionally fractionated (>25) courses of intensity-modulated radiation therapy (IMRT) (Bortfeld, et al. 2002). With hypofractionated courses, studies have shown the necessity of using multiple arcs to yield the averaging benefit to SBRT lung with VMAT (Court, et al. 2010 and Ong, et al. 2011).

Eliminating concern of the interplay effect, forward-planned, dynamic conformal arc (DCA) treatments are known to be efficient and clinically effective (Ross, et al. 2011 and Ku, et al. 2016). The contrast between DCA and VMAT is illustrated in Figure 1. Note that the VMAT plan in this figure has some MLC leaves occluding the target whereas the leaves are conformed to the shape of the target for the DCA plan. Dosimetric accuracy has also been demonstrated to decrease with increased modulation complexity and average leaf travel (Masi, et al 2013). With the high degree of precision required for hypofractionated treatments, it is important to ensure radiotherapy plans are adequately but not overly complex.

The current study reviews the DCA planning method within the RayStation (RaySearch Laboratories, Stockholm, Sweden) treatment planning system and compares DCA to VMAT plans, highlighting the appropriate selection of each technique. We also evaluate RayStation's capability to generate segment-weight optimized dynamic conformal arc (SWO-DCA) plans, where a non-uniform number of monitor units (MUs) as a function of the gantry angle is determined by inverse planning techniques. This type of optimization is, similar to optimization for VMAT, driven towards fulfilment of a set of user-defined objective functions. However, only the number of MUs per segment are varied during the optimization, the MLC leaves are kept unchanged at their conformed positions relative to the target volume.

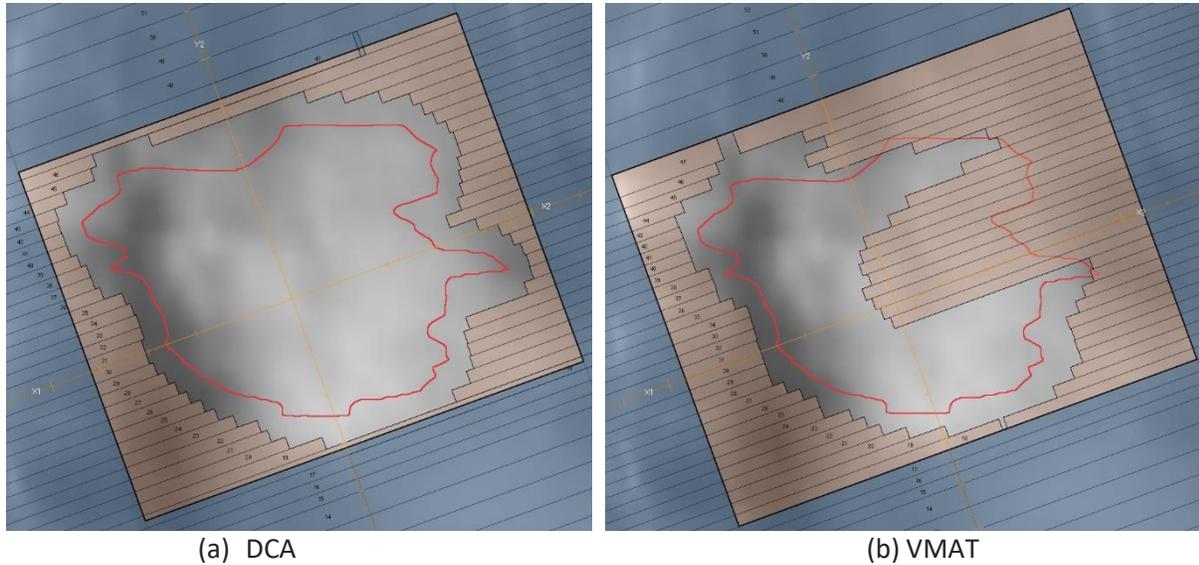
(a) DCA  (b) VMAT

**Figure 1.** Typical segment shapes for DCA and VMAT treating a target volume indicated by the red contour. MLC leaves are shown in brown and jaws shown in blue.

## 2. METHODS
Treatment plans were generated for three anonymized patients (patients 1-3) using RayStation 7.0. Treatment planning was performed with the average image derived from all phases of a four-dimensional computed tomography (4DCT) dataset. An ITV was created using the maximum intensity projection 4DCT. Three plans were developed per patient: a standard DCA plan, an SWO-DCA plan, and a VMAT plan. All using a single coplanar arc with a 6 MV energy beam from a TrueBeam linear accelerator (Varian Medical Systems, Palo Alto, CA). Arc lengths of 255, 210, and 240 degrees were used for patients 1, 2, and 3, respectively. The fractionation schedule used was 60 Gy in 8 fractions. The DCA plan of each patient was created by conforming the MLC to the ITV with a margin of 0.5 cm and scaling the number of MUs to achieve D95% to the prescription level. The SWO-DCA and VMAT plans were generated by optimization with respect to objectives defined in accordance with patient-specific clinical goals, as summarized in Tables 2-4. The standard DCA plan was used as initial point for the SWO-DCA optimizations.

Treatment plans were evaluated with respect to level of clinical goal fulfilment and dose-volume histograms (DVHs). Treatment plan complexity was assessed in term of the modulation complexity score (MCS), the score being introduced by McNiven, et al. (2010) for static-field IMRT and later adapted to VMAT by Masi, et al. (2013). The MCS score depends on the leaf position variability between adjacent active leaves and the aperture area variability. The score is dimensionless and ranges from 0 to 1, with a value of 1 corresponding to the lowest possible complexity (a rectangular field). The set of active leaf pairs used for the MCS evaluation was defined as the leaf pairs with a tip gap inside the jaw opening that is greater than the minimum dynamic tip for at least one control point. Plan complexity was also assessed in terms of total leaf travel, averaged over the active leaf pairs, and total MU variability.

## 3. RESULTS
The dose distributions of the treatment plans are illustrated with DVHs in Figure 2 and 3D dose distributions for a transversal slice in Figures 3-5. The contours for regions of interest (ROIs) in these figures are indicated in colors in accordance with Figure 2. The examined plan complexity metrics are summarized in Table 1 and the level of clinical goal fulfilment per patient case summarized in Tables 2-4.

In the first patient case, all clinical goals were achieved with the three treatment techniques (Table 2). In the second patient case, clinical goals were achieved, with exception of the spinal cord with the standard DCA plan (Table 3). The third case had the most complex geometry, with a large centrally located target proximal the spinal cord. In this case, neither DCA nor SWO-DCA plans could be created to fulfill all clinical goals (Table 4). The standard DCA plan failed on sparing of the spinal cord and trachea, whereas the SWO-DCA plan failed the spinal cord. In addition to improved sparing of the trachea, SWO-DCA also led to improved sparing of the esophagus compared to standard DCA, as evident in Figure 2(c). The VMAT plan fulfilled all goals for patient 3 (Table 4).

Figures 3-5 show standard DCA plans produce relatively symmetrical dose distributions. The SWO-DCA and VMAT plans, in contrast, yielded heavily weighted anterior and posterior dose delivery for all three cases. The VMAT plans generally exhibited a higher level of complexity than standard DCA and SWO-DCA plans according to evaluated complexity metrics. The exception to this general pattern was the MCS value for patient 2, which was lower for SWO-DCA than VMAT. The lower MCS value (higher complexity) of the SWO-DCA plan was due to a higher level of aperture area variability, which for this patient was caused by irregular target geometry. The VMAT plan is, arguably, the more complex plan having a factor 2.8 higher average leaf travel per degree and a factor 2.3 higher MU variability per degree.

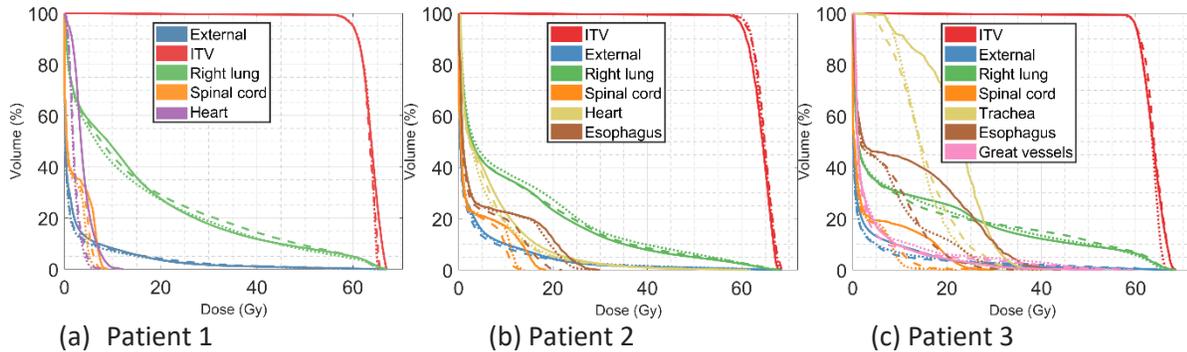

(a) Patient 1    (b) Patient 2    (c) Patient 3

**Figure 2.** DVHs for the three patient cases. The standard DCA plan is indicated by solid lines, the SWO-DCA plan indicated by dashed lines, and the VMAT plan indicated by dotted lines.

**Table 1.** Summary of obtained plan complexity metrics per patient and plan generation technique. MCS = modulation complexity score. LT = average leaf travel. ΔMU = MU variation per degree.

| Patient | Metric | DCA | SWO-DCA | VMAT |
| --- | --- | --- | --- | --- |
| P1 | MCS (-) | 0.62 | 0.63 | 0.60 |
|  | LT (cm) | 2.5 | 2.5 | 4.6 |
|  | ΔMU (MU/°) | 0.0 | 25.5 | 34.9 |
| P2 | MCS (-) | 0.45 | 0.43 | 0.45 |
|  | LT (cm) | 5.1 | 5.1 | 14.6 |
|  | ΔMU (MU/°) | 0.0 | 60.9 | 138.0 |
| P3 | MCS (-) | 0.51 | 0.50 | 0.37 |
|  | LT (cm) | 2.9 | 2.9 | 21.0 |
|  | ΔMU (MU/°) | 0.0 | 103.3 | 129.1 |

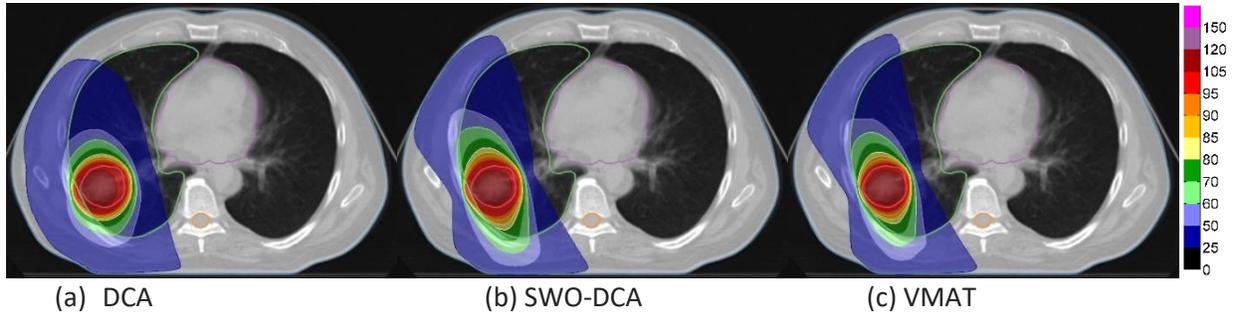

(a) DCA            (b) SWO-DCA           (c) VMAT

**Figure 3.** Dose distributions for patient 1 for a transversal cut through the isocenter, overlaid on the planning CT. Contours for ROIs are indicated by solid lines according to the color scheme of Figure 2(a). The color table is in percent of the prescription dose (60 Gy).

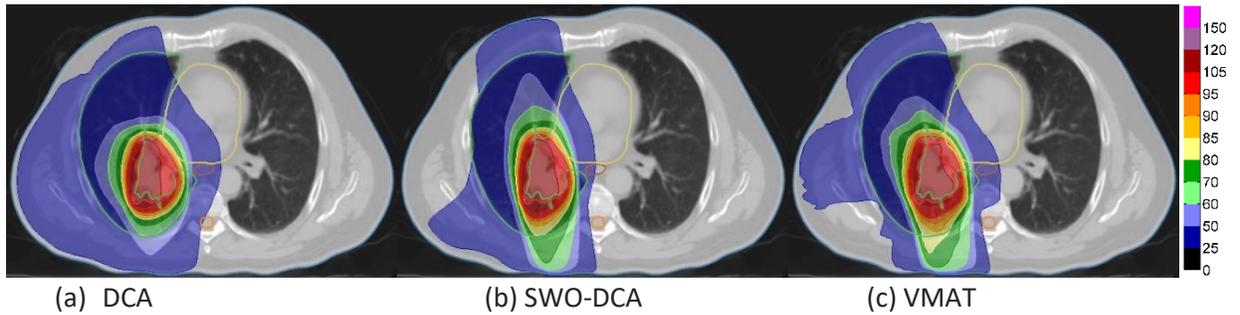

(a) DCA            (b) SWO-DCA           (c) VMAT

**Figure 4.** Dose distributions for patient 2 for a transversal cut through the isocenter, overlaid on the planning CT. Contours for ROIs are indicated by solid lines according to the color scheme of Figure 2(b). The color table is in percent of the prescription dose (60 Gy).

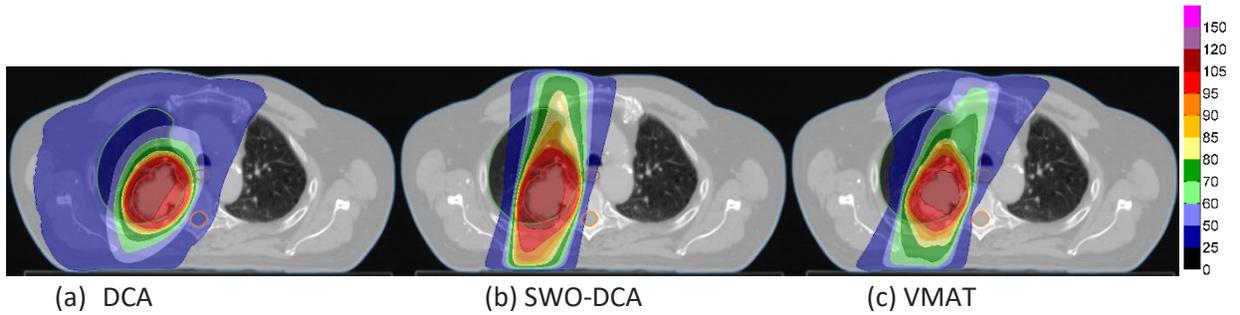

(a) DCA            (b) SWO-DCA           (c) VMAT

**Figure 5.** Dose distributions for patient 3 for a transversal cut through the isocenter, overlaid on the planning CT. Contours for ROIs are indicated by solid lines according to the color scheme of Figure 3(c). The color table is in percent of the prescription dose (60 Gy).

**Table 2.** Fulfilment of clinical goals on maximum dose at volume for Patient 1. Satisfied goals are indicated in green and violated goals indicated in red.

| ROI | Maximum dose at volume | Resulting dose (Gy) at volume | | |
|---|---|---|---|---|
| | | DCA | SWO-DCA | VMAT |
| Spinal cord | 13.5 Gy at 0.5 cm$^3$ | 7.93 | 6.96 | 6.16 |
| | 22.5 Gy at 0.25 cm$^3$ | 8.07 | 7.19 | 6.43 |
| | 30.0 Gy at 0 cm$^3$ | 9.01 | 8.19 | 7.40 |
| Lung (Right & Left) | 12.5 Gy at 1500 cm$^3$ | 2.71 | 1.78 | 1.59 |
| | 13.5 Gy at 1000 cm$^3$ | 8.74 | 7.40 | 6.36 |
| Heart | 32.0 Gy at 15 cm$^3$ | 8.87 | 5.45 | 4.90 |
| | 63.0 Gy at 0 cm$^3$ | 12.2 | 8.39 | 7.17 |
| Trachea | 18.0 Gy at 4 cm$^3$ | 1.25 | 0.87 | 0.80 |
| | 63.0 Gy at 0 cm$^3$ | 12.0 | 6.89 | 6.48 |

**Table 3.** Fulfilment of clinical goals on maximum dose at volume for Patient 2. Satisfied goals are indicated in green and violated goals indicated in red.

| ROI | Maximum dose at volume | Resulting dose (Gy) at volume | | |
|---|---|---|---|---|
| | | DCA | SWO-DCA | VMAT |
| Spinal cord | 13.5 Gy at 0.5 cm$^3$ | 16.7 | 11.7 | 12.6 |
| | 22.5 Gy at 0.25 cm$^3$ | 17.0 | 12.4 | 13.1 |
| | 30.0 Gy at 0 cm$^3$ | 18.6 | 13.3 | 13.9 |
| Lung (Right & Left) | 12.5 Gy at 1500 cm$^3$ | 5.53 | 4.16 | 5.31 |
| | 13.5 Gy at 1000 cm$^3$ | 9.07 | 7.56 | 9.84 |
| Heart | 32.0 Gy at 15 cm$^3$ | 30.0 | 28.5 | 29.0 |
| | 63.0 Gy at 0 cm$^3$ | 61.3 | 61.3 | 62.3 |
| Esophagus | 27.5 Gy at 5 cm$^3$ | 20.8 | 14.4 | 19.4 |
| | 63.0 Gy at 0 cm$^3$ | 30.0 | 21.9 | 28.7 |

**Table 4**. Fulfilment of clinical goals on maximum dose at volume for Patient 3. Satisfied goals are indicated in green and violated goals indicated in red.

| ROI | Maximum dose at volume | Resulting dose (Gy) at volume | | |
|---|---|---|---|---|
| | | DCA | SWO-DCA | VMAT |
| Spinal cord | 13.5 Gy at 0.5 cm$^3$ | 25.0 | 17.2 | 12.7 |
| | 22.5 Gy at 0.25 cm$^3$ | 25.9 | 18.8 | 14.6 |
| | 30.0 Gy at 0 cm$^3$ | 30.0 | 29.8 | 22.5 |
| Lung (Right & Left) | 12.5 Gy at 1500 cm$^3$ | 11.3 | 0.89 | 0.96 |
| | 13.5 Gy at 1000 cm$^3$ | 44.3 | 2.31 | 2.46 |
| Heart | 32.0 Gy at 15 cm$^3$ | 0.65 | 0.61 | 0.65 |
| | 63.0 Gy at 0 cm$^3$ | 0.87 | 0.78 | 0.84 |
| Trachea | 18.0 Gy at 4 cm$^3$ | 25.8 | 17.9 | 16.5 |
| | 63.0 Gy at 0 cm$^3$ | 40.2 | 40.3 | 38.6 |
| Esophagus | 27.5 Gy at 5 cm$^3$ | 24.6 | 12.5 | 11.5 |
| | 63.0 Gy at 0 cm$^3$ | 44.1 | 36.0 | 40.0 |
| Great vessels | 47.0 Gy at 10 cm$^3$ | 17.2 | 19.5 | 25.2 |
| | 63.0 Gy at 0 cm$^3$ | 52.7 | 59.0 | 59.9 |

## 4. DISCUSSION

Our results demonstrate DCA plans can achieve satisfactory and nearly equivalent plans to VMAT under favorable conditions. Based on the cases examined in this study, situations when DCA can provide dose distributions of comparable quality to VMAT are lesions that are not proximal to dose-limiting OARs. We observed that complex cases with proximal OARs are better served with advanced treatment techniques such as VMAT.

It was observed that segment weight optimization can considerably improve DCA plan quality with negligible change in plan complexity. Dose distributions of the SWO-DCA plans were also observed to be similar to VMAT plans. Thus, we have demonstrated that DCA, and particularly SWO-DCA, is a simple technique to create single arc lung SBRT plans of comparable quality with VMAT, while eliminating concerns of interplay and reducing complexity.

**ACKNOWLEDGEMENTS**
The authors thank Kjell Eriksson for constructive comments on the study's experimental design and Cameron Ditty for early work on DCA in RayStation.